# Superfluid Dynamics of a Bose-Einstein Condensate in a one-dimensional Optical Super-Lattice

Aranya B Bhattacherjee
Department of Physics, Atma Ram Sanatan Dharma College, University of Delhi (South Campus), Dhaula Kuan, New Delhi-110021, India
E.Mail: abhattac@bol.net.in

**Abstract**: We derive and study the Bloch and Bogoliubov spectrum of a Bose-Einstein condensate (BEC) confined in a one-dimensional optical superlattice (created by interference between a primary optical lattice and a secondary optical lattice of small strength), using the Bogoliubov approximation and the hydrodynamic theory with mode coupling. We show that a BEC in an optical superlattice experiences two different tunneling parameters and hence behaves like a chain of diatomic lattice. We derive expressions for the tunneling parameters as a function of the strength of the primary and secondary lattice. This gives rise to a gapped branch in addition to the gapless acoustical branch in the Bogoliubov spectrum. The spectrum strongly depends on the strength of the secondary lattice, the interaction parameter and the number density of atoms. The effective mass is found to increase as the depth of the secondary optical lattice increases, a property that was utilized in ref. [20] to achieve the Tonks-Girardeau regime. The coupling between the inhomogeneous density in the radial plane and the density modulation along the optical lattice gives rise to multibranch Bogoliubov spectrum.

## 1. Introduction

Ultracold bosons trapped in an optical lattice have been widely used recently as a model system for the study of some fundamental concepts of quantum physics. Such optical potentials have been used to study Josephson effects [1], squeezed states [2], Landau-Zener tunneling and Bloch oscillations [3], the classical [4] and quantum [5] superfluid – Mott insulator transition. Bose-Einstein condensate (BEC) in optical lattice offer a close analogy with solid-state physics as atoms evolve in an almost defect free optical potential like the electrons in an ion matrix. An important and promising application under study is quantum computation in optical lattices [6]. Optical lattices are, therefore, of particular interest from the perspective of both fundamental quantum physics and its connection to applications.

An interesting extension of optical lattices is optical superlattices. Optical superlattices are generated experimentally by the superposition of optical lattices with different periods [7]. The light shifted potential of the superlattice is denoted as

$$V(z) = V_1 \cos^2\left(\frac{\pi z}{d_1}\right) + V_2 \cos^2\left(\frac{\pi z}{d_2} + \varphi\right) \qquad (1)$$

Here $d_1$ and $d_2 > d_1$ are respectively, the primary and secondary lattice potentials. $V_1$ and $V_2$ are the respective amplitudes. The secondary lattice acts as a perturbation and hence $V_2 < V_1$. $\varphi$ is the phase of the secondary lattice. Theoretical interest in optical superlattice started only recently. These include work on fractional filling Mott insulator domains [8], dark [9] and gap [10] solitons, the Mott-Peierls transition [11], non-mean field effects [12] an phase diagram of BEC in two color superlattices [13]. In a recent interesting work, the analogue of the optical branch in solid-state physics was predicted in an optical superlattice [14]. The purpose of this paper is to study some superfluid properties of a cigar shaped Bose-Einstein condensate trapped in a one–dimensional optical superlattice. We will explicitly discuss the change in the behaviour of the system as a function of the depth of the secondary lattice. First in section 2 we give a systematic derivation of the Josephson coupling parameters and the Bose-Hubbard Hamiltonian of the cold atoms in an optical super-lattice. We then derive and study the Bloch spectrum in section 3 and the Bogoliubov spectrum using the Bogoliubov approximation in section 4 and the hydrodynamic approach with mode coupling in section 5. Expression for the Bloch energy, effective mass, Bogoluibov amplitudes and quadrapole modes in an optical superlattice are new results of this paper.

## 2. The Energy Functional and the Bose-Hubbard Hamiltonian

In this section we derive the Bose-Hubbard Hamiltonian of the optical superlattice following the works of refs. [15] and [16]. In the classical (mean-field) approximation, the BEC dynamics at T=0 is governed by the Gross-Pitaevskii energy functional [15]:

$$E_0 = \int dV \psi^+(r,z)\left[-\frac{\hbar^2}{2M}\nabla^2 + V_{ho}(r,z) + V_{op}(z) + \frac{U}{2}|\psi(r,z)|^2\right]\psi(r,z) \qquad (2)$$

Here, $V_{ho}(r,z) = \frac{M}{2}(\omega_r^2 r^2 + \omega_z^2 z^2)$ is the harmonic trap potential and $V_{op}(z) = E_R\left(s_1 \cos^2\left(\frac{\pi z}{d}\right) + s_2 \cos^2\left(\frac{\pi z}{2d}\right)\right)$ is the optical superlattice potential. We have taken a particular case of $d_2 = 2d_1 = 2d$. $s_1$ and $s_2$ are the dimensionless amplitudes of the primary and the secondary superlattice potentials with $s_1 > s_2$. $E_R = \hbar^2 \pi^2 / 2Md^2$ is the recoil energy ($\omega_R = E_R/\hbar$ is the corresponding recoil energy) of the primary lattice. $U = 4\pi a \hbar^2 / M$ is the strength of the two-body interaction energy and $a$ is the two-body scattering length. We take $\omega_r \gg \omega_z$ so that an elongate cigar shaped BEC is formed. The frequency corresponding to the minima of the optical superlattice is $\omega_s \approx \sqrt{s_1} \hbar \pi^2 / Md^2$.

The BEC is initially loaded into the primary lattice and the secondary lattice is switched on slowly. The frequency of each minima of the primary lattice is not perturbed significantly by the addition of the secondary lattice. $\omega_s \gg \omega_z$ so that the optical lattice dominates over the harmonic potential along the z-direction and hence the harmonic potential is neglected. The strong laser intensity will give rise to an array of several quasi-two-dimensional pancake shaped condensates. Because of the quantum tunneling, the overlap between the wave functions between two consecutive layers can be sufficient to ensure full coherence. The 3-dmensional wave function of the condensate is written as

$$\psi(r,z) = \sum_j \psi_j(r) W(z - z_j) \qquad (3)$$

Here, $\psi_j(r)$ is the wavefunction of the condensate along the radial direction at the site $j$ and $W(z - z_j)$ is the localized wavefunction at the $j$ site along the z-direction. $W(z - z_j)$ is written as [15]

$$W(z - z_j) = \left(\frac{M\omega_s}{\pi \hbar}\right)^{1/4} \exp\left(-\frac{M\omega_s}{2\hbar}(z - z_j)^2\right) \qquad (4)$$

Here, $z_j = jd$. Sustituting eqn.(3) into eqn.(2) and considering only the nearest-neighbor interactions, we get the following energy functional

$$E_0 = \sum_j \int dx dy \left[-\frac{\hbar^2}{2M} \psi_j^+ \nabla_r^2 \psi_j + V_{ho}(x,y)|\psi_j|^2\right] + \frac{U_{eff}}{2} \sum_j \int dx dy\, \psi_j^+ \psi_j^+ \psi_j \psi_j$$
$$- \sum_j J_j \int dx dy \left[\psi_{j\pm 1}^+ \psi_j + \psi_j^+ \psi_{j\pm 1}\right]$$

(5)

Here $J_j$ is the site dependent strength of the Josephson coupling and is different when going from $j-1$ to $j$ and $j$ to $j+1$.

$$J_j = -\int dz W(z)\left[-\frac{\hbar^2}{2M}\nabla_z^2 + V_{op}(z)\right]W(z+d) \tag{6}$$

In order to understand the two different types of Josephson coupling, we refer to fig.1, where we have plotted the primary, secondary and the combined two optical potential. The BEC is initially placed into the primary optical lattice and then the secondary lattice is switched on slowly creating the effective potential as shown in the figure. It is clear from the figure that the BEC encounters an additional barrier from the secondary lattice when going from j to j+1. On the other hand the BEC encounters a potential well of the secondary lattice when going from j-1 site to j site. One can show using eqn.(6) that there are distinctly, two Josephson coupling parameters $J_0 \pm \Delta_0/2$ where

$$J_0 \approx \left(\frac{2}{\pi}\right)^{3/2} E_R \left[\left(\frac{\pi^2}{2}-2\right)s_1 - s_2 - \sqrt{s_1}\right]\exp-\left(\frac{\pi^2\sqrt{s_1}}{4}\right) \tag{7}$$

$$\Delta_0 \approx 2\left(\frac{2}{\pi}\right)^{3/2} E_R s_2 \exp-\left(\frac{\pi^2\sqrt{s_1}}{4}\right) \tag{8}$$

Equations (7) and (8) indicate that as the strength of the secondary lattice $s_2$ increases, the increase in $\Delta_0$ is twice as large as the decrease in $J_0$. Consequently, one of the barrier corresponding to the tunneling $J_0 + \Delta_0/2$ decreases while the other barrier corresponding to $J_0 - \Delta_0/2$ increases. The strength of the effective on-site interaction energy is $U_{eff} = U\int dz|W(z)|^4$.

The Bose-Hubbard Hamiltonian corresponding to the energy functional of eqn.(5) is similar to an effective 1D Bose-Hubbard Hamiltonian in which each lattice site is replaced by a layer with radial confinement.

$$H = -\sum_j J_j \left[a_j^+ a_{j+1} + a_{j+1}^+ a_j\right] + \frac{U'_{eff}}{2} a_j^+ a_j^+ a_j a_j \tag{9}$$

Here $U'_{eff} = \frac{U_{eff}}{V_{2D}}$, $V_{2D}$ is the 2D area of radial confinement and $J_j = \left(J_0 - \frac{\Delta_0}{2}(-1)^j\right)$.

The Bose-Hubbard Hamiltonian is an appropriate model when the loading process produces atoms in the lowest vibrational state of each well, with a chemical potential smaller than the distance to the first vibrationally excited state.

## 3. The Bloch Chemical Potential and Bloch Energy

In this section and the next, we consider mean-field solutions of eqn.(9). In the mean field approximation, the operators $a_j$ and $a_j^+$ are classical c-numbers, $a_j = \varphi_j$. Stationary states with a fixed total number of particles (N) are obtained by requiring that the variation of $H - \mu N$ with respect to $\varphi_j^*$ vanish. $\mu$ is the chemical potential. This yields the eigenvalue equation

$$U'_{eff}|\varphi_j|^2 \varphi_j - J_j \varphi_{j+1} - J_{j-1}\varphi_{j-1} - \mu\varphi_j = 0 \qquad (10)$$

We write $\varphi_j$ as

$$\varphi_j = g_k \exp(ikj2d) = \left(u_k + i(-1)^j v_k\right)\exp(ikj2d) \qquad (11)$$

Note that the periodicity of the system is now that of the secondary lattice. We assume that each site of the lattice is perfectly equivalent due to the symmetries of the system so that the population is same at each site. This is particularly true if we neglect the harmonic trap frequency along the direction of the optical lattice. The solution of equation (10) for $k = 0$ gives the ground state of the system. This state corresponds to a condensate at rest in the frame of the optical lattice. Instead the solutions of equation (10) with $k \neq 0$, describes same single-particle wavefunction, move together with respect to the optical lattice-giving rise to a constant current. Experimentally, such states can be created by turning on adiabatically the intensity of the primary lattice moving with a fixed velocity. We take the 2D average density of atoms per site of the lattice as $n_0 = N/IV_{2D}$. $N$ is the total number of atoms and $I$ is the total number of sites in the one dimensional optical lattice, with $|g_k|^2 = N/I$ the number of atoms per site. We thus get the lowest eigenvalues as

$$\mu(k,n_0) = U_{eff} n_0 - \sqrt{4J_0^2 \cos^2(2kd) + \Delta_0^2 \sin^2(2kd)} \qquad (12)$$

The eigenvalues of eqn.(12) corresponds to the chemical potential for $k=0$. The energy per particle defined as

$$\varepsilon(k) = \frac{1}{n_0}\int \mu(k,n_0)dn_0 \qquad (13)$$

$$\varepsilon(k) = \frac{U_{eff} n_0}{2} - \sqrt{4J_0^2 \cos^2(2kd) + \Delta_0^2 \sin^2(2kd)} \qquad (14)$$

The second term in equation (14) is the tight binding expression for the energy of a Bloch state for a single particle in an optical superlattice of equation (1). The energy per particle $\varepsilon(k)$ of stationary Bloch-wave configuration consists of the motion of the whole condensate and carries a current constant in time and uniform in space (Bloch bands). The effective mass of the atoms in the optical superlattice is given as

$$\frac{1}{m^*} = \frac{1}{\hbar^2} \frac{\partial^2 \varepsilon}{\partial k^2}\bigg|_{k=0} = \frac{2d^2(4J_0^2 - \Delta_0^2)}{J_0 \hbar^2} \qquad (15)$$

The effective mass is found to increase with increasing $\Delta_0$. An increase in $\Delta_0$ implies a decrease in tunneling between wells corresponding to $\frac{\pi z}{d} = 1.5$ and $\frac{\pi z}{d} = 2.5$ of fig.1 and a corresponding increase in tunneling between wells corresponding to $\frac{\pi z}{d} = 0.5$ $\frac{\pi z}{d} = 1.5$. The atoms get trapped in an effective potential well consisting of two lattice sites (e.g. $\frac{\pi z}{d} = 0.5$ and $\frac{\pi z}{d} = 1.5$ in fig. 1). Figure 2 shows a plot of eqn. (14) as a function of $kd$ for $J_0/E_R = 0.5$ and $\Delta_0/E_R = 0.1$ (dashed line) and $\Delta_0/E_R = 0.4$ (solid line). The ground state energy has been subtracted in fig.2. The effect of $\Delta_0$ is clearly visible near the band edge where the curve bends. The single particle energy decreases with increasing $\Delta_0$. The effect is more pronounced near the band edge.

**4. The Bogoliubov Dispersion relation**

The Bogoliubov spectrum of elementary excitation describes the energy of small perturbations with quasi-momentum $q$ on top of a macroscopically populated state with quasi-momentum $k$. In the Bogoliubov approximation [17], we write the annihilation operator in terms of the c-number part and a fluctuation operator as

$$a_j = (\varphi_j + \delta_j) \exp\left(-i\mu t/\hbar\right) \qquad (16)$$

The resulting Bogoliubov equations for the fluctuation operator $\delta_j$ in the optical superlattice take the following form:

$$i\hbar \dot{\delta}_j = (2U_{eff} n_0 - \mu)\delta_j - J_j \delta_{j+1} - J_{j-1}\delta_{j-1} + U_{eff} n_0 \delta_j^+ \qquad (17)$$

We are interested in the ground state solutions with $k = 0$. The above equation is solved by constructing quasi-particles for the lattice, which diagonalize the Hamiltonian i.e.

$$\delta_j = \frac{1}{\sqrt{I}} \sum_q \left\{ u_j^q \alpha_q \exp i(qj2d - \omega_q t) - v_j^{q*} \alpha_q^+ \exp -i(qj2d - \omega_q t) \right\} \tag{18}$$

The quasi-particles obey the usual Bose-commutation relations

$$[\alpha_q, \alpha_{q'}^+] = \delta_{qq'} \tag{19}$$

The excitation amplitudes obey the following periodic boundary conditions:

$$u_{j+1}^q = u_{j-1}^q, \quad v_{j+1}^q = v_{j-1}^q \tag{20}$$

Finally the phonon excitation frequencies are found to be

$$\hbar \omega_{q,\pm} = \sqrt{\tilde{\varepsilon}_{q,\pm}(2n_0 U_{eff} + \tilde{\varepsilon}_{q,\pm})} \tag{21a}$$

$$\tilde{\varepsilon}_{q,\pm} = 2J_0 \pm \sqrt{4J_0^2 \cos^2(2qd) + \Delta_0^2 \sin(2qd)} \tag{21b}$$

The $\omega_{q,-}$ branch is the familiar acoustical branch and it can be checked by putting $\Delta_0 = 0$ in eqn.(21) and we get back the familiar dispersion relation in an uniform optical lattice with periodicity $2d$ [18]. The second branch $\omega_{q,+}$ is a new branch, which was reported recently [14] and it is the analogue of the optical branch in solid-state physics. We call it the gapped branch (note that it is not correct to call this branch as the optical branch since the frequency of this branch does not lie in the optical region). Fig.3 shows a plot of eqn.(21) for $\Delta_0/E_R = 0.1$, $J_0/E_R = 0.4$ and $n_0 U_{eff}/E_R = 0.3$. A gap is clearly visible at the band edge ($qd = \pi/4$). The acoustical branch of Eqn.(21a) has a form similar to the well known Bogoliubov spectrum of uniform gases, the energy $\varepsilon_{q,-}$ replacing the free particle energy $\hbar^2 q^2/2m$. In the long wavelength limit ($k \to 0$), the $\omega_{q,-}$ branch is approximately written as

$$\omega_{q,-} = q \sqrt{\frac{n_0 U_{eff}}{m^*}} \tag{22}$$

The spectrum is sound like and resembling the Bogoluibov spectrum of uniform gas with effective mass $m^*$. The sound velocity is seen to be $\sqrt{\frac{n_0 U_{eff}}{m^*}}$.

On the other hand the $\omega_{q,+}$ branch in the low wavelength limit is approximated as

$$\omega_{q,+} \approx \frac{1}{\hbar}\left(\sqrt{8J_0(2J_0 + n_0 U_{eff})} - \frac{\hbar^2 q^2}{2m^*}\frac{(4J_0 + n_0 U_{eff})}{\sqrt{8J_0(2J_0 + n_0 U_{eff})}}\right) \qquad (23)$$

The Bogoliubov amplitudes are derived as

$$|u_j^q|^2 = |u_{j+1}^q|^2 = \frac{1}{2}\left(\frac{\tilde{\varepsilon}_{q,-} + n_0 U_{eff} + \hbar\omega_{q,-}}{\hbar\omega_{q,-}}\right) \qquad (24)$$

$$|v_j^q|^2 = |v_{j+1}^q|^2 = \frac{1}{2}\left(\frac{\tilde{\varepsilon}_{q,-} + n_0 U_{eff} - \hbar\omega_{q,-}}{\hbar\omega_{q,-}}\right) \qquad (25)$$

The effect of $\Delta_0$ on the Bogoliubov amplitudes is appreciable only near the band edge $kd = \pi/4$.

### 5. Multibranch Bogoliubov Spectrum

In the previous section, we had calculated the excitation frequencies of the Bogoliubov quasi-particles without taking into account the coupling between the inhomogeneous density in the radial plane and the density modulation along the optical lattice. In actual experiments, the mode couplings are always present and should change the Bogoliubov spectrum obtained in the previous section. When we excite the BEC along the direction of the optical lattice, other low energy transverse modes are also excited and these transverse modes modify the axial modes. Huang and Wu [14] had calculated the spectrum of the phonon and monopole modes of a BEC in an optical superlattice by considering only the coupling between the phonon and the breathing modes. But actually the acoustical mode is coupled not only to the breathing mode but also with other low energy modes having zero angular momentum [15]. The calculations in this section following ref. [15] takes this mode coupling into account. Mode coupling is known to reduce the spectrum strongly and it should be taken into account for calculating the spectrum correctly [15]. We use the discrete hydrodynamic approach to recalculate the Bogoliubov spectrum in the presence of mode coupling. Our starting point now is the energy functional of eqn.(5). The Heisenberg equation of motion for the Bosonic order parameter is

$$i\hbar\dot{\psi}_j = \left[-\frac{\hbar^2}{2M}\nabla_r^2 + V_{ho} + U_{eff}\psi_j^+\psi_j\right]\psi_j - \left(\psi_{j-1}J_{j-1} + \psi_{j+1}J_j\right) \qquad (26)$$

here $J_j = J_0 - \frac{\Delta_0}{2}(-1)^j$.

Writing $\psi_j = \sqrt{n_j}\exp(i\theta_j)$, where $n_j$ is the Bosonic density at the site $j$ and $\theta_j$ is the corresponding phase, we get the following equations of motion for $n_j$ and $\theta_j$ after neglecting the quantum pressure term

$$\dot{n}_j = -\frac{\hbar}{M}\vec{\nabla}_r \cdot (n_j \nabla_r \theta_j) + \frac{2}{\hbar}\{J_{j-1}\sqrt{n_j n_{j-1}}\sin(\theta_j - \theta_{j-1}) - J_j\sqrt{n_j n_{j+1}}\sin(\theta_{j+1} - \theta_j)\} \tag{27}$$

$$\hbar\dot{\theta}_j = -\frac{\hbar^2}{2M}(\nabla_r \cdot \theta_j)^2 + \left\{J_{j-1}\sqrt{\frac{n_{j-1}}{n_j}}\cos(\theta_j - \theta_{j-1}) + J_j\sqrt{\frac{n_{j+1}}{n_j}}\cos(\theta_{j+1} - \theta_j)\right\} - V_{ho} - U_{eff} n_j \tag{28}$$

In the limit of small oscillations, we linearize eqns.(27) and (28) around the equilibrium state $n_j = n_0 + \delta n_j$ and $\theta_j = \delta\theta_j$. $n_0$ is the equilibrium mean 2D condensate density at each site and $\delta n_j$ and $\delta\theta_j$ are the small deviations in density and phase. The equations of motion for the density and phase fluctuations are written as

$$\delta\dot{n}_j = -\frac{\hbar}{M}\nabla_r \cdot [n_0(r)\nabla_r \delta\theta_j] + \frac{2n_0}{\hbar}\left[2J_0\delta\theta_j - \left(J_0 - \frac{\Delta_0}{2}(-1)^{j-1}\right)\delta\theta_{j-1} - \left(J_0 - \frac{\Delta_0}{2}(-1)^j\right)\delta\theta_{j+1}\right] \tag{29}$$

$$\hbar\delta\dot{\theta}_j = -U_{eff}\delta n_j - \frac{1}{2n_0(r)}\left[2J_0\delta n_j - \left(J_0 - \frac{\Delta_0}{2}(-1)^{j-1}\right)\delta n_{j-1} - \left(J_0 - \frac{\Delta_0}{2}(-1)^j\right)\delta n_{j+1}\right] \tag{30}$$

From eqns.(29) and (30), we get the second order equation of motion for the density fluctuations as

$$\delta\ddot{n}_j = \frac{2\delta n_j}{\hbar^2}\left[-2U_{eff}n_0 J_0 - 3J_0^2 - \frac{\Delta_0^2}{4}\right] + \frac{2\delta n_{j-1}}{\hbar^2}(2J_0 + n_0 U_{eff})\left[J_0 - \frac{\Delta_0}{2}\right] + \frac{2\delta n_{j+1}}{\hbar^2}(2J_0 + n_0 U_{eff})\left[J_0 + \frac{\Delta_0}{2}\right] - \frac{(J_0^2 - \Delta_0^2/4)}{\hbar^2}[\delta n_{j+2} + \delta n_{j-2}] \tag{31}$$

In the limit of small oscillations the solutions of eqn.(31) have the form

$$\delta n_j = (\delta n_a(r) + i(-1)^j \delta n_b(r))\exp i(jq2d - \omega_l t) \tag{32}$$

here $\delta n_a$ and $\delta n_b$ are the amplitudes. Eqn. (31) shows that there are two distinct types of amplitude fluctuations. $\delta n_j = (\delta n_a(r) + i\delta n_b(r))\exp i(jq2d - \omega_l t)$ for j even and $\delta n_j = (\delta n_a(r) - i\delta n_b(r))\exp i(jq2d - \omega_l t)$ for j odd. $q$ is the quasi-momentum of the small perturbations. The quasi-momentum $k$ is put to zero here. The two amplitudes satisfies the following two coupled hydrodynamic equations

$$-\omega_l^2 \delta n_a = \frac{U_{eff}}{M}\vec{\nabla}_r \cdot (n_0 \nabla_r \delta n_a) + \frac{\delta n_a}{\hbar^2}\varepsilon_- - \frac{2\Delta_0}{\hbar^2}(2J_0 + U_{eff} n_0)\sin(2qd)\delta n_b \tag{33}$$

$$\omega_l^2 \delta n_b = -\frac{U_{eff}}{M}\vec{\nabla}_r \cdot (n_0 \nabla_r \delta n_b) + \frac{\delta n_b}{\hbar^2}\varepsilon_+ + \frac{2\Delta_0}{\hbar^2}(2J_0 + U_{eff} n_0)\sin(2qd)\delta n_a \tag{34}$$

Where

$$\varepsilon_\pm = (4J_0(2J_0 + U_{eff} n_0)\cos(2qd) \pm (4J_0^2 - \Delta_0^2)\cos^2(2qd) \pm 4J_0(J_0 + n_0 U_{eff}) \pm \Delta_0^2) \tag{34a}$$

Here $l$ is a set of two quantum numbers: radial quantum number ($n_r$) and the angular quantum number ($m$). The solutions of eqns.(33) and (34) can be obtained by expanding the density fluctuations $\delta n_a(r)$ and $\delta n_b(r)$ as

$$\delta n_a(r) = \sum_l b_{l,a} \delta n_l(r,\phi), \quad \delta n_b(r) = \sum_l b_{l,b} \delta n_l(r,\phi) \tag{35}$$

$\delta n_l(r,\phi)$ is the normalized eigenfunction for $q = 0$ [15,19].

$$\delta n_l(r,\phi) = \frac{(1 + 2n_r + |m|)^{1/2}}{(\pi R_0^2)^{1/2}} \tilde{r}^{|m|} P_{n_r}^{(|m|,0)}(1 - 2\tilde{r}^2)\exp(im\phi) \tag{36}$$

here $P_n^{(a,b)}(x)$ is the Jacobi polynomial of order $n$. $\tilde{r} = r/R_0$, $R_0$ is the radius of the condensate at each site. $R_0 = 2\mu/M\omega_R^2$, $\omega_R$ is the recoil frequency. $\phi$ is the polar angle.

In the above expansion, we assume that the radial and angular momentum of the condensate at each site is the same and is unaffected by the asymmetric tunneling parameter. Substituting eqn.(35) into eqns.(33) and (34), we obtain the following coupled eigenvalues equations

$$-\omega_l^2 b_{l,a} = -(|m| + 2n_r(n_r + |m| + 1))b_{l,a} + \frac{\left(b_{l,a} - \sum_{l'} M_{l,l'} b_{l',a}\right)}{\hbar^2}\varepsilon_-$$
$$-\frac{2\Delta_0}{\hbar^2}(2J_0 + U_{eff} n_0)\sin(2qd)\left(b_{l,b} - \sum_{l'} M_{l,l'} b_{l',b}\right)$$
(37)

$$\omega_l^2 b_{l,b} = -(|m| + 2n_r(n_r + |m| + 1))b_{l,b} + \frac{\left(b_{l,b} - \sum_{l'} M_{l,l'} b_{l',b}\right)}{\hbar^2}\varepsilon_+$$
$$+\frac{2\Delta_0}{\hbar^2}(2J_0 + U_{eff} n_0)\sin(2qd)\left(b_{l,a} - \sum_{l'} M_{l,l'} b_{l',a}\right)$$
(38)

The matrix element $M_{ll'}$ is given by

$$M_{ll'} = \frac{(1 + 2n_r + |m|)}{\pi}\int d^2\tilde{r}\,\tilde{r}^{2+|m|+|m'|} P_{n_r}^{(|m|,0)}(1 - 2\tilde{r}^2)\exp(i(m - m')\phi)P_{n_{r'}}^{(|m'|,0)}(1 - 2\tilde{r}^2)$$
(39)

The coupled equations (37) and (38) can be solved numerically for any values of $l$ and $l'$ but in order to appreciate the difference between these results and those obtained in the previous section, we derive analytical results for the lowest two modes ($m = 0$, $n_r = 0,1$). Solving the corresponding eigenvalues equations we get the following Multibranch Bogoliubov spectrum

$$\tilde{\omega}_{n_r}^2 = 2n_r(n_r + 1) + (1 - M_{n_r,n_r})\tilde{\omega}_{q,\pm}^2$$
(40)

here $\tilde{\omega}_{n_r}^2 = \omega_{n_r}^2/\omega_R^2$ and $\tilde{\omega}_{q,\pm}^2 = \omega_{q,\pm}^2/\omega_R^2$

In the limit of long wavelength, the $n_r = 0$ mode is phonon like with a sound velocity $c = \sqrt{n_0 U_{eff}/2m^*}$ which is smaller by a factor $\sqrt{2}$ with respect to the sound velocity obtained in the previous section. $m^* = J_0\hbar^2/2d^2(4J_0^2 - \Delta_0^2)$ is the effective mass of the atoms in the optical superlattice. The sound velocity decreases as the strength of the secondary lattice increases. Figure 4 shows the lowest four modes for $\Delta_0/E_R = 0.1$, $J_0/E_R = 0.4$ and $n_0 U_{eff}/E_R = 0.2$. The $n_r = 0, l = 0$ modes (the lower two curves) are the same as that in figure 3. Clearly, the lowest two modes found from this approach are $\sqrt{2}$ times smaller than those derived in the previous section by the Bogoluibov approach because of the average over the radial variable. The lowest branch (solid curve) corresponds to the Bogoliubov axial acoustical mode with no radial nodes. The next branch (dashed curve) is the gapped axial branch with no radial nodes. The third branch corresponds to the acoustical branch with one radial node (in phase breathing mode) while the fourth branch (dashed) corresponds to the gapped branch with one radial node (out of phase breathing mode). In the long wavelength limit ($q \to 0$), the Bogoluibov modes in the m=0 sector can be approximated as

$$\omega_i^2 = \Omega_i^2 + \alpha_i^2 q^2 \tag{41}$$

where $i$ =AB (acoustical branch), GB (gapped branch), IB (in phase breathing), OB (out of phase breathing).

$$\Omega_{AB}^2 = 0, \ \alpha_{AB}^2 = \frac{n_0 U_{eff}}{2m^*} = c_{AB}^2 \ \text{(velocity of sound for the acoustical mode)} \tag{42a}$$

$$\Omega_{GB}^2 = \frac{4J_0(2J_0 + n_0 U_{eff})}{\hbar^2}, \ \alpha_{GB}^2 = -\frac{(4J_0 + n_0 U_{eff})}{2m^*} \tag{42b}$$

$$\Omega_{IB}^2 = 4\omega_r^2, \ \alpha_{IB}^2 = \frac{n_0 U_{eff}}{2m^*} \tag{42c}$$

$$\Omega_{OB}^2 = 4\omega_r^2 + \frac{4J_0(2J_0 + n_0 U_{eff})}{\hbar^2}, \ \alpha_{OB}^2 = -\frac{(4J_0 + n_0 U_{eff})}{2m^*} \tag{42d}$$

The lowest energy quadrapole modes ($n_r = 0, m \pm 2$) can also be calculated easily and these modes strongly depend on other quadrapole modes with radial nodes ($n_r \neq 0, m \pm 2$). Neglecting the couplings among all other modes in the $m = \pm 2$ sector by setting $l' = (n_r, 2)$ in Eqns. (33)-(35), we easily get the following spectrum

$$\tilde{\omega}_{n_r}^2 = 2 + 2n_r(n_r + 3) + (1 - M_{n_r,2,n_r,2})\tilde{\omega}_{q,\pm}^2 \tag{43}$$

The quadrapole spectrum of the Eqns.(43) is a new result of this paper and lowest energy ($n_r = 0, m \pm 2$) in phase (IQ) and out of phase (OQ) quadrapole modes are plotted in fig.5. In the high wavelength regime, the quadrapole modes of fig.5 are written as

$$\omega_{IQ}^2 = 2\omega_r^2 + \frac{3n_0 U_{eff}}{4m^*}q^2 \qquad (44a)$$

$$\omega_{OQ}^2 = 2\omega_r^2 + \frac{6J_0(2J_0 + n_0 U_{eff})}{\hbar^2} - \frac{3(4J_0 + n_0 U_{eff})}{4m^*}q^2 \qquad (44b)$$

It is worth noting that the $q=0$ in phase breathing (IB) and the in phase quadrapole mode are not influenced by $J_0$, $\Delta_0$ and $U_{eff}$.

The spectrum of elementary excitations derived above can be measured by inelastic Bragg scattering experiments where the system is exposed to a weak external moving optical potential along the z-direction, transferring momentum $p_z(\omega, q)$ and energy $\hbar\omega$. The multibranch Bogoliubov spectrum is obtained by measuring the locations of the peak in $p_z(\omega, q)$ for different $q$.

## 6. Conclusions

In conclusion, we have studied both the Bloch and Bogoliubov excitation spectrum of the axial quasiparticles of a Bose-Einstein Condensate confined in a one-dimensional optical superlattice with two different tunneling parameters. Two different approaches were adopted, namely the Bogoliubov approximation and the hydrodynamic approach taking into account the coupling with the radial modes. The physics of this system with two tunneling parameters was found to be equivalent to that of a diatomic chain modified by the inter-atomic interactions. In particular, we found a gapped branch (analogous to the optical branch) in addition to the familiar acoustical gapless branch. The coupling to the radial modes generates a rich spectrum with many modes of higher energy. The approach presented in this work generates the full spectrum of the BEC in an optical superlattice by taking into account various mode coupling. Mode coupling reduces the Bogoliubov spectrum significantly. In the tight binding regime, analytic expressions for the Bogoliubov amplitudes have also been reported. The energy gap (energy difference between the gapless branch and the first gapped branch) at the edge of the Brillouin zone depends on the strength of the secondary optical lattice. The gap increases as the strength of the secondary lattice increases. This gap is also found to increase with increasing inter-atomic interaction. The Bloch energy is found to decrease near the band edge with increasing $\Delta_0$. We have shown that the effective mass is increased and the sound velocity decreased by the presence of the secondary lattice. A similar technique has been used earlier [20] to achieve the Tonks-Girardeau regime.

### Acknowledgments

I acknowledge support by the Deutsche Forschungsgemeinschaft within the SFB Transregio 12. Thanks to Professor R. Graham, for providing me the facilities for this work at the University of Düisberg-Essen, Germany. Also I am grateful to K.Krutitsky for useful discussions.

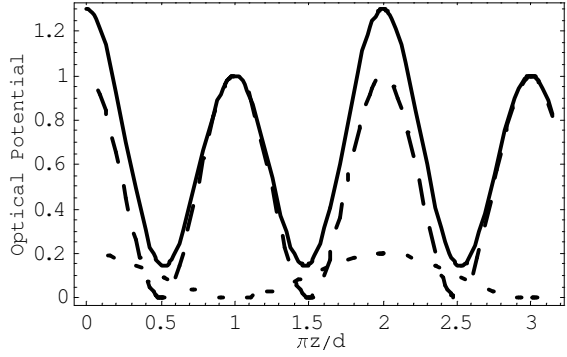

**Figure 1**: A plot showing the primary lattice (long dashed line), the secondary lattice (short dashed line) and the superlattice (solid line) created as a result as a superposition of the primary and the secondary lattice. The BEC encounters an additional barrier from the secondary lattice when going from j ($\frac{\pi z}{d}=1.5$) to j+1 ($\frac{\pi z}{d}=2.5$). On the other hand the BEC encounters a potential well of the secondary lattice when going from j-1 ($\frac{\pi z}{d}=0.5$) site to j site.

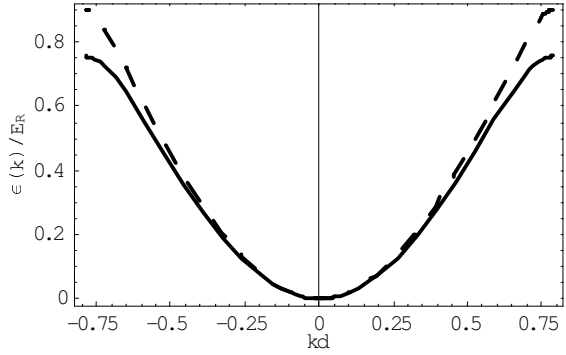

**Figure 2**: Plot of Bloch energy as a function of *kd* for $J_0/E_R=0.5$ and $\Delta_0/E_R=0.1$ (dashed line) and $\Delta_0/E_R=0.4$ (solid line). The ground sate energy has been subtracted.

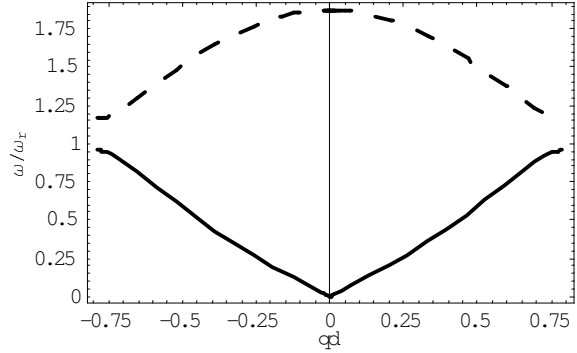

**Figure 3**: Lowest two Bogoluibov bands for $\Delta_0/E_R = 0.1$, $J_0/E_R = 0.4$ and $n_0 U_{eff}/E_R = 0.2$. The solid line is the gapless acoustical branch while the dashed line is the gapped branch. A gap is clearly visible at the band edge ($qd = \pi/2$).

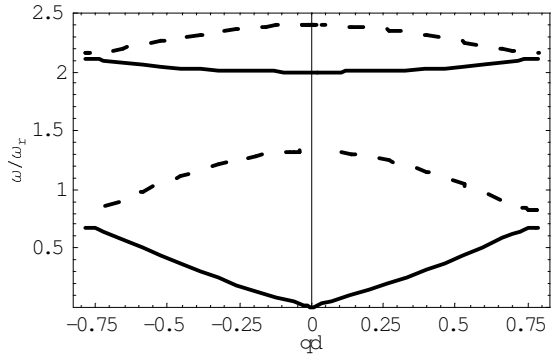

**Figure 4**: Lowest four Bogoluibov modes for $\Delta_0/E_R = 0.1$, $J_0/E_R = 0.4$ and $n_0 U_{eff}/E_R = 0.2$. The $n_r = 0, l = 0$ modes (the lower two curves) are the same as that in figure 3. The upper two curves correspond to $n_r = 1, l = 0$. The lower solid line corresponds to the acoustical branch while the dashed line corresponds to the gapped branch. The upper solid line is the in-phase breathing mode and the upper dashed curve is the out-of phase breathing mode.

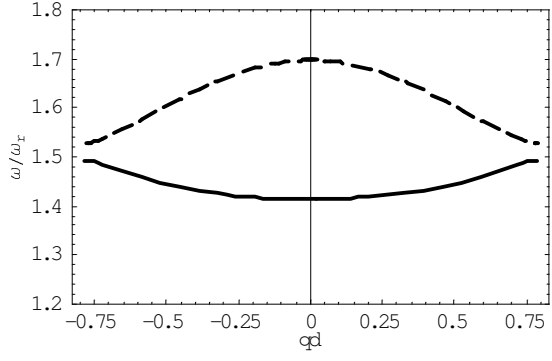

**Figure 5**: Plots of the lowest energy Bogoliubov quadrapole modes in the $m = \pm 2$ sector. Solid line corresponds to in phase mode and the dashed line corresponds to out of phase mode. Here, $\Delta_0/E_R = 0.1$, $J_0/E_R = 0.4$ and $n_0 U_{eff}/E_R = 0.2$.